# Active Interface Characteristics of Heterogeneously Integrated GaAsSb/Si Photodiodes

Manisha Muduli[1], Yongkang Xia[1], Seunghyun Lee[1**], Nathan Gajowski[1], Chris Chae[2], Siddharth Rajan[1], Jinwoo Hwang[2], Shamsul Arafin[1], Sanjay Krishna[1*]

[1]Department of Electrical and Computer Engineering, The Ohio State University, Ohio, 43201, USA
[2]Department of Material Science and Engineering, The Ohio State University, Ohio, 43201, USA

* Corresponding author: krishna.53@osu.edu
** Now at Department of Electrical Engineering, University of Texas, Arlington, 76019

There is increased interest in the heterogeneous integration of various compound semiconductors with Si for a variety of electronic and photonic applications. This paper focuses on integrating GaAsSb (with absorption in the C-band at 1550nm) with silicon to fabricate photodiodes, leveraging epitaxial layer transfer (ELT) methods. Two ELT techniques—epitaxial lift-off (ELO) and macro-transfer printing (MTP)—are compared for transferring GaAsSb films from InP substrates to Si, forming PIN diodes. Characterization through atomic force microscopy (AFM), and transmission electron microscopy (TEM) exhibits a high-quality, defect-free interface. Current-voltage (IV) measurements and capacitance-voltage (CV) analysis validate the quality and functionality of the heterostructures. Photocurrent measurements at room temperature and 200 K demonstrate the device's photo-response at 1550 nm, highlighting the presence of an active interface.

Silicon photonics technology has emerged as an important technology with applications in optical communications, long range remote sensing, and LiDAR [1, 2]. III-V materials such as InGaAs, InP, GaAsSb, InSb, InGaAs/GaAsSb superlattice (SL) can absorb photons beyond the Silicon band-edge extending to the short-wave Infrared (SWIR) and extended short-wave infrared (e-SWIR) regime while maintaining high quantum efficiencies [3]. In particular, while III-V materials have higher absorption coefficients, Silicon has very favorable impact ionization properties that leads to high avalanche gain and low excess noise factors. Integration of III-V materials with silicon can leverage the benefits of both materials yielding superior photonic devices that are capable of absorbing in the SWIR/e-SWIR range while providing high gains and bandwidth [2, 4, 5]. III-V/Si integration is a widely studied topic for several decades [4, 6, 7, 8, 9]. Hawkins et. al. have demonstrated high bandwidth heterostructure photodiodes operating at 1.55 µm, but these devices have encountered challenges due to carrier loss at the interface [2]. Therefore, a significant research gap exists in developing III-V/Si integrated photonics with an *active interface* (where the carriers must be transported across the III-V/Si interface) that effectively absorbs at 1.55 µm.

The leading III-V semiconductor material absorbing at 1.55 µm wavelengths is an InGaAs lattice matched to the InP substrate [10]. In this work, we explore the integration of Si with GaAsSb, an iso-gap direct bandgap semiconductor lattice matched to the InP substrate. Unlike InGaAs, which is the workhorse of a 1.55 µm absorber, GaAsSb exhibits a low conduction band offset with Si (<0.1eV), which favors electron transport between GaAsSb and Si and eliminates the need for a grading layer [4, 6]. These properties make GaAsSb an excellent material for integrating with Si. The eventual goal is to make infrared avalanche photodiodes (APDs) with a GaAsSb absorber and a Si multiplier. In this work, we have demonstrated a PN photodiode with an active interface using an n-type $GaAs_{0.51}Sb_{0.49}$ (henceforth GaAsSb) absorber integrated with p-type Si. Further, we studied the material characteristics and device characteristics of the heterostructure. Successful integration of GaAsSb with Si will find its applications in remote sensing, high-speed detectors, quantum computing, guided laser systems, and quantum computing [6, 11, 12].

Several methods to integrate III-V materials have been studied in the past including wafer bonding [4], epitaxial growth [13], epitaxial layer transfer [9], and micro-transfer printing [14], etc. Wafer bonding is preferred commercially for wafer-level integration [4]. However, it can impart defects such as voids and cracks in the hetero-structure interface due to the high temperature and pressure required during the bonding process [9]. Epitaxial growth of III-V materials on Si can be challenging due to the different thermal expansion coefficients and high lattice mismatch, leading to threading dislocations on device surfaces [4, 7, 15]. Unlike other techniques, epitaxial layer transfer (ELT) avoids high temperatures or pressures and relies solely on Van der Waals forces. This method enables the integration of highly mismatched materials, such as III-V semiconductors [7, 11]. ELT can yield a pristine III-V/Si heterostructure that is free from interfacial defects [9]. Additionally, depending on the process of the ELT, the source substrate used for growing the III-V material, can be recycled, lowering the effective cost of materials and device fabrication [16]. Due to the ease and processing flexibility of ELT, the source epitaxial layer can be transferred to various target substrates including PET and glass [9, 17].

Traditionally, ELT involves a source substrate, which is used to grow the desired epitaxial layer, and a target substrate, onto which this epitaxial layer is transferred. The grown epi is then detached from the host substrate by selectively wet etching and transferred to the target substrate with polymers such as polydimethylsiloxane (PDMS) [7, 9]. This paper focuses on transferring GaAsSb epitaxial layers, grown on an InP substrate, to a silicon substrate to fabricate PIN diodes. We use two different methods of epitaxial layer transfer to form the GaAsSb/Si heterostructure. Method 1 (epitaxial lift-off) involves preserving the substrate by growing an additional sacrificial layer, which is then etched to detach the GaAsSb epitaxial layer from the InP substrate before transferring. Method 2 (macro-transfer printing) involves etching away the entire InP substrate before transferring the GaAsSb layer. Both methods have benefits and drawbacks. While epitaxial lift-off preserves the substrate, it generally requires the membrane to have small holes for the etchant to reach the sacrificial layer [18]. The holes are usually spaced closely together, which restricts the formation of large-sized devices. Alternatively, when transferring devices using this method, it requires the devices to have anchors and tethers to hold the epi floating in the etchant [19]. The tethers usually transfer along with the device to the target substrate and require additional steps for removal. While macro-transfer printing allows for the transfer of devices of any size and shape, it involves removing the entire substrate before transferring, which increases the cost of the process. The details of the two methods are discussed later.

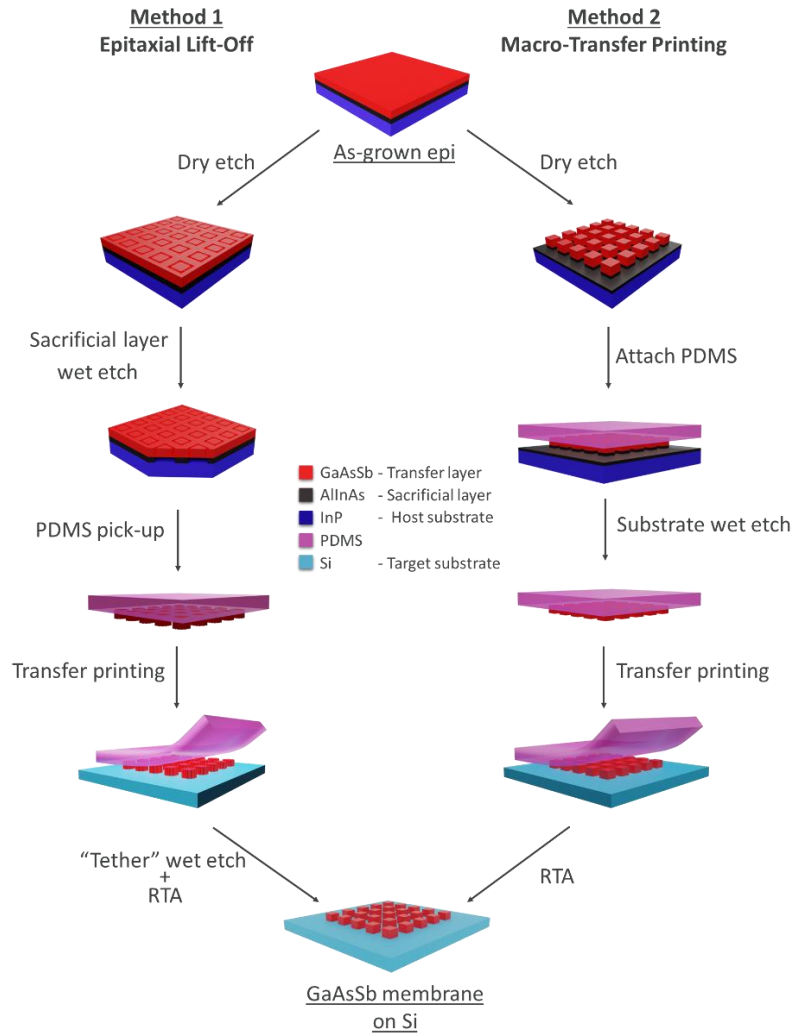

*Figure 1 Schematic illustrations of the two methods of transfer printing: (1) Epitaxial Lift-Off and (2) Macro-Transfer Printing. Epitaxial Lift-Off only etches the sacrificial layer of AlInAs before transfer, and Macro-Transfer Printing etches the InP substrate and AlInAs sacrificial layer before transfer.*

Growth of GaAsSb epitaxial films: The n++-GaAsSb/n+-GaAsSb/UID–GaAsSb/UID–AlInAs layers with a 1500 nm UID-GaAsSb were grown on an n+-InP (100) substrate using molecular beam epitaxy system. The details of the stack are shown in Table 1. The n++ and n+ GaAsSb layers were doped using Si to nominal concentrations of $1 \cdot 10^{19}\ cm^{-3}$ and $2 \cdot 10^{18}\ cm^{-3}$, respectively. The sacrificial layer of AlInAs was chosen for its etch selectivity against GaAsSb when using specific wet etchants. It was also used as a buffer layer before growing GaAsSb on the InP substrate. A coupled x-ray diffraction (XRD) measurement (Supplementary Figure 1) of the epitaxial film was used to determine the GaAsSb film had a lattice mismatch $\Delta a/a$ of $1.4 \cdot 10^{-3}$, displaying a high-quality growth of the epitaxial film.

| Layer | Material | Doping | Thickness |
|---|---|---|---|
| Nanomembrane for transfer | GaAsSb | n$^{++}$ | 20 nm |
| | GaAsSb | n$^{+}$ | 100 nm |
| | GaAsSb | UID | 1500 nm |
| Sacrificial Layer | AlInAs | UID | 350 nm |
| Substrate | InP | Semi-insulating | 500 µm |

*Table 1 Shows the epitaxially grown GaAsSb on an InP substrate.*

Fabrication of GaAsSb/Si diodes: The fabrication of the GaAsSb/Si diodes was completed by heterogeneously integrating GaAsSb films onto p-type Si substrate, to form a P-I-N diode. The epitaxial film transfer was completed using two different methods. In Method 1 (Figure 1), we etch the sacrificial AlInAs layer, enabling the GaAsSb film to suspend and tethered above InP. Subsequently, we utilize a Polydimethylsiloxane (PDMS) stamp for the pick-up, transfer, and printing onto a silicon substrate. In Method 2 (Figure 1), we attach the GaAsSb/AlInAs/InP stack onto a PDMS stamp. Then, we fully etch the InP substrate and AlInAs layers before transferring the GaAsSb film onto a Si substrate. The PDMS stamp in both methods was unpatterned and fabricated using a silicon elastomer base mixed with a curing agent at a ratio of 4:1. The details of the transfer methods are given below.

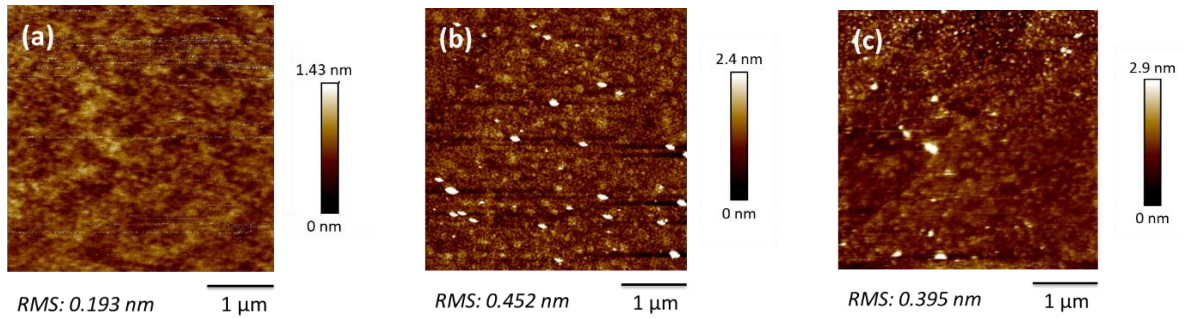

*Figure 2 AFM images of (a) as-grown GaAsSb epi on top of InP substrate, (b) GaAsSb nanomembrane on Si substrate after epitaxial lift-off, and (c) GaAsSb nanomembrane on Si substrate after macro-transfer printing. The AFM images of the samples were taken over an area of 4 µm x 4 µm. The white regions in the images denote the crests or the highest points on the film, and the dark or black spots denote the troughs or the lowest points on the film.*

Method 1 (Epitaxial Lift-Off): The GaAsSb stack was patterned using photolithography to form the desired pattern. The pattern consisted of square devices of size 300 µm x 300 µm, that were connected to the anchors using tethers. A mixture of BCl$_3$ and Ar plasma was employed for dry etching the stack down to the AlInAs layer, to facilitate the wet etching of the sacrificial layer. AlInAs is then selectively etched for approximately 30 minutes using a mixture of 36% HCl and deionized (DI) water (3:1). HCl/DI etchant etches AlInAs and InP but does not affect GaAsSb. A layer of 200 nm of SiO$_2$ is grown on the InP substrate, to protect it during the wet etching process. After the AlInAs is completely etched, the stack is washed with DI water and dried using an N$_2$ gun. Adjacently, the Silicon substrate is cleaned and prepared using 49% HF and deionized (DI) water. A PDMS stamp is then used to pick up the GaAsSb membrane and printed onto a Si substrate to form the GaAsSb/Si heterostructure. After the transfer, the GaAsSb/Si heterostructure is annealed at 300°C for 5 minutes. The devices were then patterned and etched to removed tethers, such that the size of the final square devices is 250 μm x 250 μm.

Method 2 (Macro-Transfer printing): The GaAsSb stack was patterned using photolithography to form a pattern consisting of square devices of size 300 μm x 300 μm. A mixture of $BCl_3$ and Ar plasma was employed for dry etching the stack down to the InP layer. A PDMS stamp was then attached to the epitaxial stack, such that the GaAsSb film was towards the PDMS, and the InP substrate was exposed. The PDMS-GaAsSb/AlInAs/InP stack was then dipped in an etchant solution of 37% HCl and DI water (3:1) with the addition of Sodium dodecyl sulfate (SDS) surfactant. The addition of SDS surfactant reduced the surface tension of hydrogen bubbles generated during the InP etching process, thereby minimizing the risk of PDMS-GaAsSb delamination. Once the InP and AlInAs layers are etched, the PDMS-GaAsSb stack is washed with DI water and dried using $N_2$ gun. The silicon substrate is cleaned and prepared using 49% HF and DI water like the previous method. The GaAsSb membrane is then quickly printed onto the Si substrate, without any further processing. After the transfer, the GaAsSb/Si heterostructure is annealed at 300°C for 5 minutes.

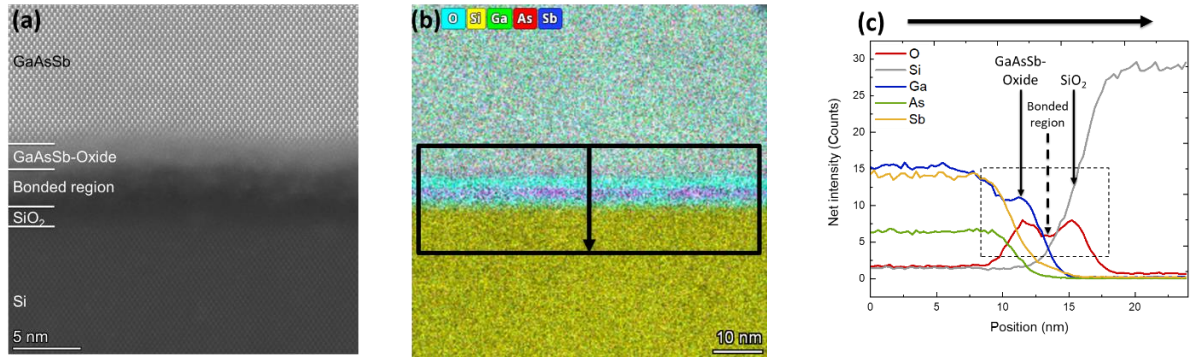

*Figure 3 (a) TEM image of the GaAsSb-Si interface, with GaAsSb as the on-zone, after GaAsSb is transferred to Si using macro-transfer printing. (d) EDS area scan of the interface depicting the elements present at the interface (c) EDS line scan of the interface depicting the formation of gallium oxide and silicon dioxide at the interface.*

In both the mentioned methods, the transfer-printing mainly depends on the differential adhesion between the PDMS-GaAsSb epitaxial layer and the GaAsSb epitaxial layer-Si substrate [20]. Once the GaAsSb membrane is printed onto the Si substrate to form the P-I-N heterostructure, the devices are patterned to deposit contact metal. A stack of Pd/Ti/Pt/Au with a thickness of 12nm/9nm/9nm/100nm respectively was deposited on GaAsSb to form the top contact. An image of the fabricated device is shown in Figure 1. A stack of Ni/Au with a thickness of 10nm/100nm was deposited on the $p^+$-Si to form the bottom contact.

The microscope image (Figure 1) of the GaAsSb/Si diode shows a uniform GaAsSb film on top of the Si substrate without any visible defects such as voids (or air gaps), depicting a high-quality epitaxial transfer. An atomic force microscopy (AFM) is performed before and after transfer to assess the effect of the transfer process on the epitaxial film quality. The obtained AFM images are shown in Figure 2 (a) - (c). AFM analysis confirmed that the wet etching and transfer printing process has minimal effect on the film quality. The AFM for as-grown GaAsSb-on-InP showed a root-mean-square (RMS) roughness of 0.193 nm. The AFM for the GaAsSb membrane transferred using Method 1 and Method 2 showed an RMS roughness of 0.452 nm and 0.395 nm respectively. The RMS roughness of the membranes indicates that the quality of the film deteriorates only slightly post-transfer. The degradation in the film quality may be due to the formation of nano-voids (airgaps) or oxides formed during the transfer process, which can be seen as white spots in the AFM images.

For heterogeneous integration, the interface between GaAsSb membrane and Si substrate is utmost important. Through epitaxial layer transfer, GaAsSb membrane adheres to the Si substrate with the help of Van der Waal forces [9]. To test the interface quality, transmission electron microcopy (TEM) along with Energy-dispersive X-ray Spectroscopy (EDS) was performed on an GaAsSb/Si sample transferred using macro-transfer printing. The TEM and EDS images for the same are shown in Figure 3 (a) – (c). The TEM and EDS analysis of the sample transferred using epitaxial lift-off are shown in Supplementary Figure 2. The TEM image in Figure 3 (a) depicts a high-quality interface, free of voids, but it also reveals the presence of native oxide that is formed during the transfer process. The total oxide thickness is measured around 6 nm, with 1.6 nm of silicon dioxide, 1.9 nm of gallium oxide, and 2.6 nm of bonded region, which is an amorphous mixture of silicon dioxide and gallium dioxide. Further, the EDS area scan is shown in Figure 3 (b) and the EDS line scan is shown in Figure 3 (c), which confirms the formation of gallium oxide formed on the side of GaAsSb film, and silicon dioxide formed on the side of Si substrate. The bonded region has the presence of gallium, silicon, and oxygen, which depicts that the oxide from the two surfaces bonded as an amorphous mixture to form a good interface.

$$V_{oc} = \frac{KT}{q} \ln\left(\frac{J_{ph}}{J_{ss}} + 1\right) \qquad \text{... Equation (1)}$$

Where
- $V_{oc}$ = Open circuit voltage
- $K$ = Boltzmann constant
- $T$ = Temperature
- $q$ = Charge
- $J_{ph}$ = Photocurrent density
- $J_{ss}$ = Saturation Current density

The epitaxial layer transfer was followed by metal deposition for contact formation. The devices were then tested for current-voltage (IV) characteristics. Figure 4 (a) displays the dark current characteristics of a device fabricated using macro-transfer printing (device A) and a device fabricated using epitaxial lift-off (device B). The rectification in the curve shows the formation of PIN junction in the GaAsSb/Si heterostructure. The dark current density obtained from devices fabricated with the two different methods was measured to be similar. The obtained ideality factors are 1.86 ± 0.5 for device A and 1.58 ± 0.3 for device B. The ideality factor for these devices was relatively high, and there was a significant error margin across multiple devices, due to the high series resistance and high contact resistance. Although on the higher side, the ideality factor for both devices falls in the expected range, indicating that the recombination in the heterostructure is limited by both carrier types. The series resistance was calculated to be 0.45 kΩ and 1.39 kΩ for devices A and B respectively. The high series resistance is primarily due to the metal contact on the $n^+$-GaAsSb, with the contact resistivity calculated to be 56 mΩ-cm$^2$. The high contact resistivity arises from Fermi-level pinning at the valence band, which complicates achieving low contact resistivity in $n^+$-GaAsSb [21]. To reduce the contact resistivity, further optimization of the metal stack is necessary. Additionally, the on-off ratio at ± 0.5 V of device A and device B were calculated as 510 and 812 respectively, which indicates a good quality heterostructure photodiode. Since the dark current characteristics of both devices were similar, Device A was chosen for all the further characterization.

The turn-on voltage for Device A was calculated to be 0.54 V using linear extrapolation of Figure 4 (b), with a rectification ratio of $2 \cdot 10^4$, indicating a reliable and efficient rectifying diode. Device A was subsequently tested for photocurrent at wavelength 1550 nm, both at room temperature and at a cooled temperature of 200 K (Figure 4 (c)). Device A exhibited a slight photocurrent at room temperature; the dark current density at -1 V was measured to be $5.5 \cdot 10^{-5} \, A/cm^2$, and the photocurrent density at the same voltage was measured to be $8.8 \cdot 10^{-5} \, A/cm^2$. The open circuit voltage ($V_{oc}$) was measured to be 20 mV for a wavelength of 1550 nm. $V_{oc}$ is an important metric for PN and PIN photodiodes, that denotes the voltage generated due to device illumination. The expected $V_{oc}$ for the GaAsSb/Si heterostructure was calculated to be 28.4 mV at a saturation current of $2.5 \cdot 10^{-8}$ using Equation (1). The experimental and theoretical calculations showed a discrepancy of 8.4 mV for the open circuit voltage. Since the interface significantly influences the $V_{oc}$, the discrepancy can be attributed to an imperfect interface between GaAsSb and Si [22].

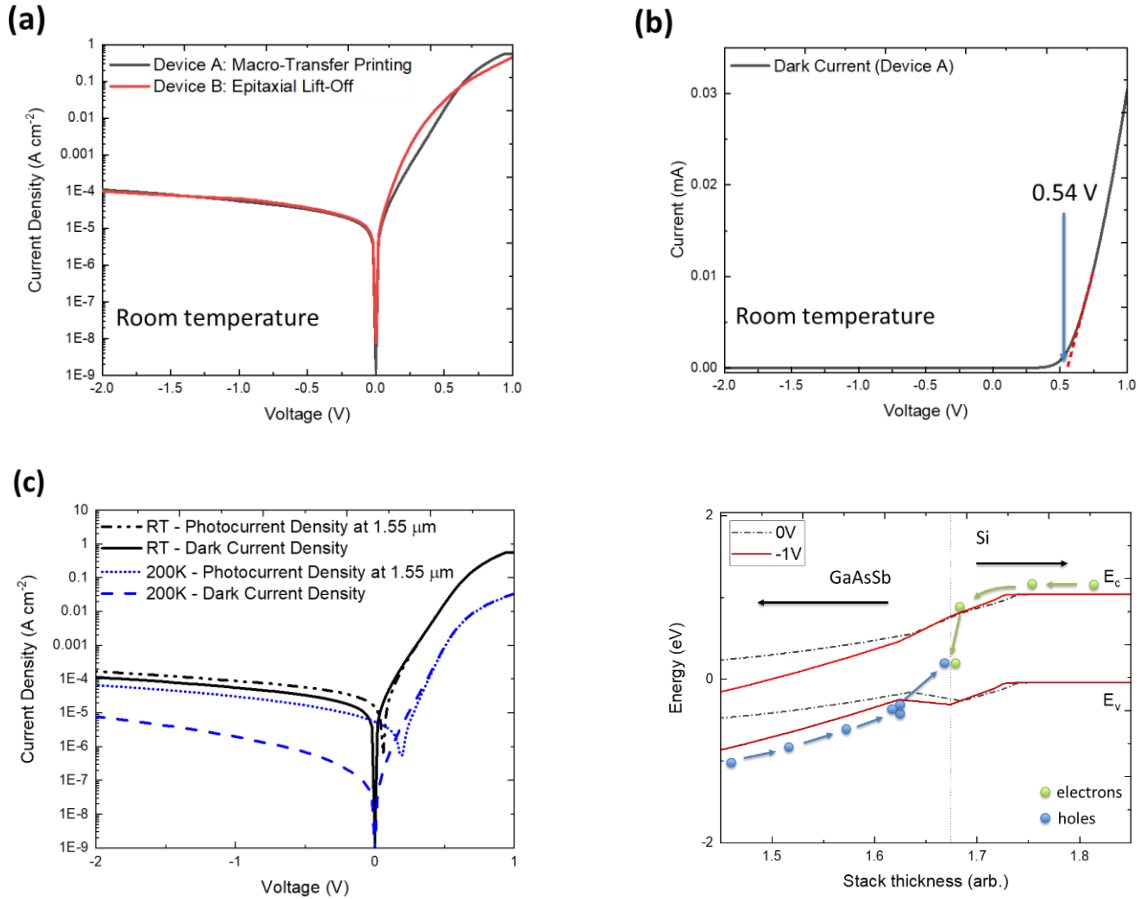

*Figure 4 Measured current-voltage characteristics of (a) devices made with epitaxial lift-off and macro-transfer printing, (b) linear scale of dark current-voltage measurements of Device A depicting the turn-on voltage of the diode, (c) logarithmic scale dark current density and photocurrent density at room temperature and at 200K of a device made with macro-transfer printing, and (d) Silvaco simulation for band-diagram of GaAsSb/Si heterostructure at the interface, for 0V bias and for -1V applied bias, depicting accumulation of carriers at the -1V applied bias.*

To confirm the photodiode response, device A was cooled to 200 K and the photocurrent was measured again. Cooling the device, showed a clear difference between the dark current density and photocurrent density, depicting the photodiode response to 1.55 µm (Figure 3(c)). The photocurrent at 1.55 indicates the presence of an active interface in the heterostructure. At the cooled temperature, the dark current density at -1 was measured to be $1.95 \cdot 10^{-5}\ A/cm^2$, and the photocurrent at the same voltage was measured to be $3.05 \cdot 10^{-5}\ A/cm^2$. The reduction in the dark current can be factored for the temperature dependence of the diffusion current and generation-recombination current of a diode [23]. Hence, we see a significant reduction in the dark current, but not much difference in the photocurrent.

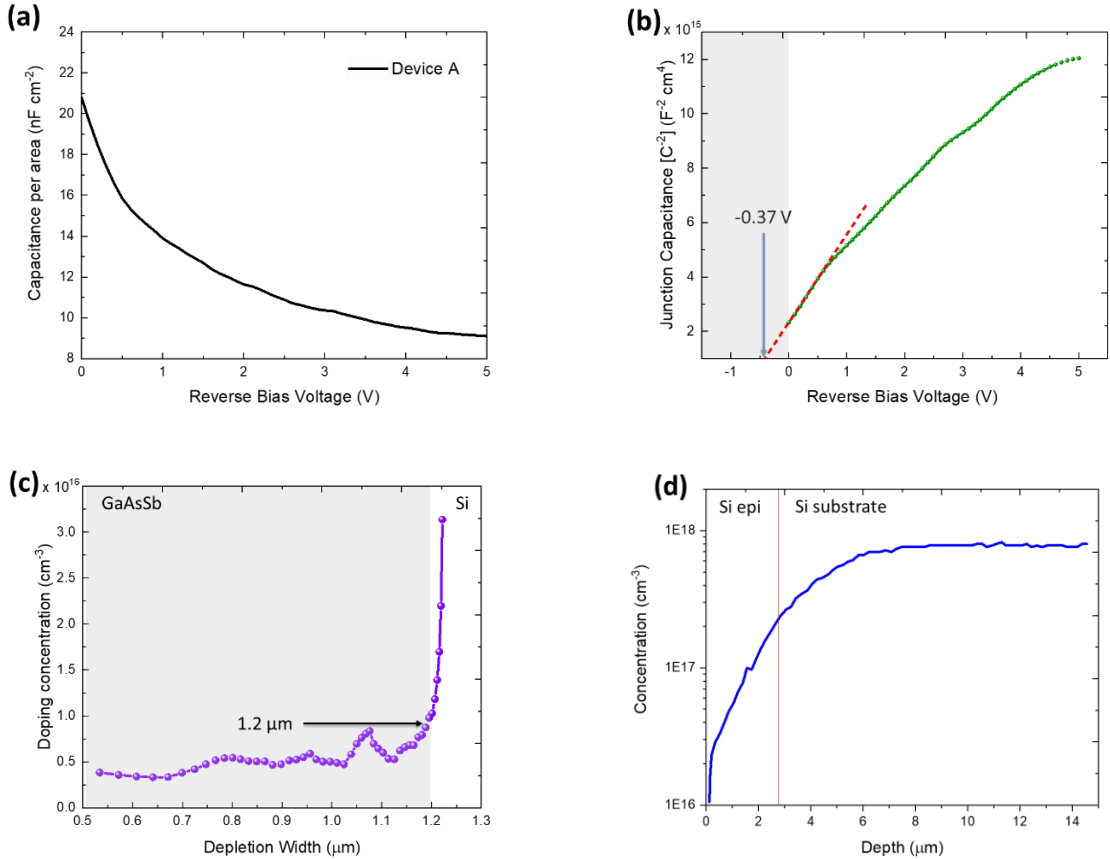

*Figure 5 (a) Measured capacitance-voltage of Device A, (b) Junction capacitance vs reverse bias of the diode depicting the built-in voltage, (c) shows doping concentration vs depletion width showing the depletion in GaAsSb layer and the depletion in Si, and (d) SIMS profile of the p-type silicon with epitaxially grown p⁻-Si (epi) on top of a p⁺-Si substrate.*

Further, the $V_{oc}$ improved to 180 mV for the photocurrent. The increase in $V_{oc}$ due to cooling can be attributed to reduced carrier recombination, improved bandgap alignment, and reduced dark current of the photodiode [24-27]. Despite the improvement in the $V_{oc}$, it is still less as compared to the InGaAs/Si heterostructures, which are direct competitors to GaAsSb/Si heterostructures [6].

A Silvaco simulation was conducted to further understand the nature of this heterostructure. Since the native oxide thickness is very less as compared to the individual stack thickness, the simulation was done for a structure without the presence of native oxides. The band-diagram near the GaAsSb/Si interface is shown in Figure 4 (d). The figure shows the band-diagram at equilibrium (0 V) and an applied bias of -1 V. It should be noted that the UID-GaAsSb is an n-doped absorber, meaning that the photo-generated majority carriers are holes, while the photo-generated minority carriers are electrons. Experimental XPS data shown in [6] and Silvaco simulation shown in Figure 4(d) suggest that the valence band offset between n-GaAsSb and p-Si is high, which could result in a hole accumulation region under a reverse bias voltage. Under Illumination, even though there are enough photo-generated carriers, this hole accumulation region may cause a reduction in the recombination between the holes and electrons, leading to reduction in the effective photo-current in the diode. A similar behavior can be observed in the GaAsSb/Si heterostructure with native oxide, and the Silvaco simulation for the same is shown in *Supplementary Figure 4*. Additionally, the native silicon dioxide not only acts as a trap center but also presents a significant barrier to carrier movement. These factors contribute to the low photocurrent as well as the low $V_{oc}$ of the heterostructure.

Figure 5 (a) shows the capacitance-voltage (CV) characteristics, conducted on Device A at 3 MHz frequency to determine the built-in voltage of the heterostructure, and to understand the depletion region in the heterostructure. The junction capacitance and doping profile are presented in Figure 5 (b) and (c). The built-in voltage of the heterostructure was determined by extrapolating from the junction capacitance graph and found to be 0.37 V (which is less than the 0.54 V turn-on voltage). For an ideal diode, the turn-on voltage equals to the built in voltage. However, due to high series resistance and contact resistance in the fabricated photodiode, the turn-on voltage exceeds the obtained built-in voltage.

The background doping for UID GaAsSb was determined to be around $5 \cdot 10^{15}\ cm^{-3}$, which suggests a good quality material growth [5]. Since the p-region of the Si is graded from $1 \cdot 10^{16}\ cm^{-3}$ to $1 \cdot 10^{18}\ cm^{-3}$ (Figure 5 (d)), calculating the depletion at zero bias is challenging. The depletion width at zero bias is extracted from Figure 5 (c) to be 586 nm, which includes the depletion in both UID-GaAsSb and Si. In Figure 4(c), it can be observed that the doping concentration remains approximately constant at $5 \cdot 10^{15}\ cm^{-3}$ up to a depletion width of 1.2 µm. Beyond this point, the doping concentration gradually increases from $1 \cdot 10^{16}\ cm^{-3}$ to $3 \cdot 10^{16}\ cm^{-3}$. This increase in doping concentration indicates the depletion in the Si substrate. A secondary ion mass spectrometry (SIMS) measurement was conducted to confirm the dopant concentration of the silicon substrate, as shown in Figure 5(d). The doping concentrations of UID GaAsSb and p-Si are similar at the heterostructure junction. Therefore, it is expected that the Si layer will begin to deplete before the UID GaAsSb layer is fully depleted, as observed in Figure 5(c).

From the IV and CV analysis, it is determined that studying the heterostructure and its interface is challenging due to several factors such as high contact resistance at the n+ GaAsSb, oxide formation at the GaAsSb/Si interface and the unusual depletion width characteristics. However, determining the interface charge is crucial for understanding whether the interface acts as a recombination center in the heterostructure. Determining the interface charge could potentially explain the low photocurrent observed at room temperature. To understand the interface further, the heterostructure will be further studied using device simulation and CV analysis using different UID thicknesses.

In conclusion, we reported the fabrication of GaAsSb/Si heterostructure using two different methods of epitaxial transfer. The results demonstrated that both methods yielded high-quality GaAsSb films on silicon substrates, with minimal degradation in film quality post-transfer, as confirmed by AFM and TEM analyses. The fabricated diodes exhibited promising IV characteristics, indicating successful formation of PIN junctions with an *active interface*. Photocurrent measurements at room temperature and 200 K depicted the photo-response of the heterostructure to 1550 nm wavelength. This photo-response at 1.55 µm demonstrates the potential of GaAsSb/Si heterostructures and represents the initial step towards developing high-speed photonic devices operating in the C-band. Further studies, including device simulation and CV analysis with varying UID thicknesses, are necessary to better understand the interface characteristics and improve device performance.

## ACKNOWLEDGEMENT

This work was supported by Intel's Center for Advanced Semiconductor Fabrication Research and Education (CAFÉ) program under Grant No. GR129846.

## DATA AVAILABILITY

The supporting data for the findings of this study are available from the corresponding author upon reasonable request.

Supplementary Information

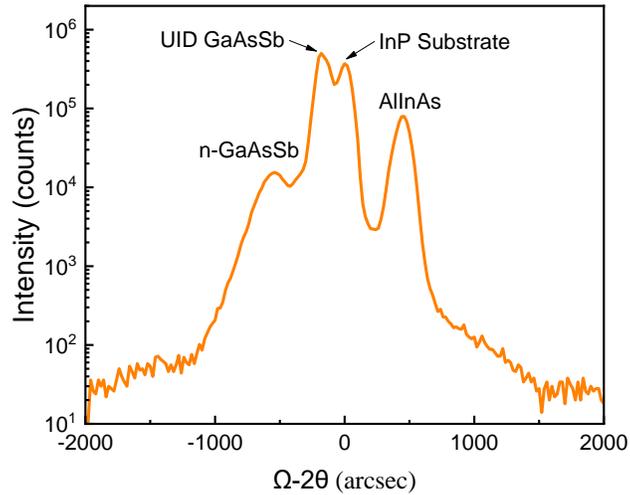

*Supplementary Figure 1 XRD measurement of the epitaxially grown GaAsSb on InP substrate shows good lattice matching between GaAsSb and InP. The peaks for n-GaAsSb, UID-GaAsSb, UID-AlInAs and InP substrate are shown in the graph.*

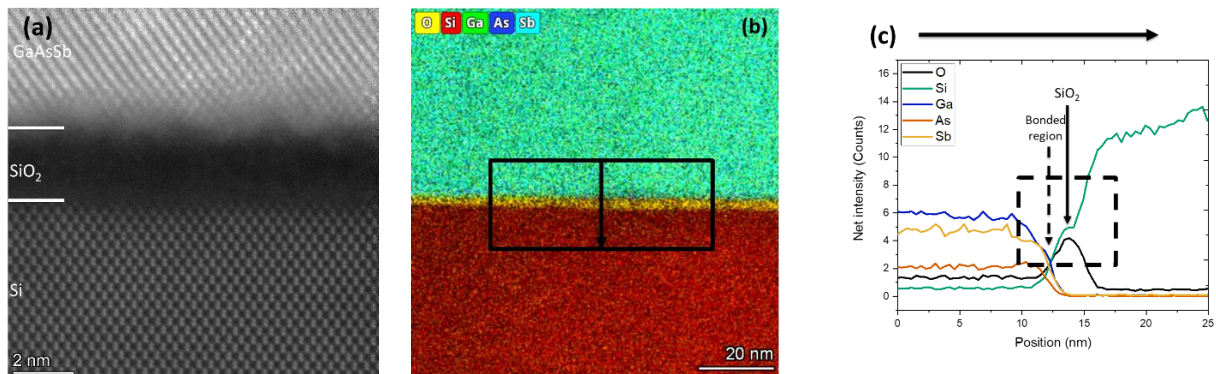

*Supplementary Figure 2 (a) TEM image of the GaAsSb-Si interface, with Si as the on-zone, after GaAsSb is transferred to Si using epitaxial lift-off. (d) EDS area scan of the interface depicting the elements presents at the interface (c) EDS line scan of the interface depicting the formation of gallium oxide and silicon di oxide at the interface.*

The TEM image shows nearly 2.3 nm Silicon dioxide and <1nm of GaAsSb-Oxide in the sample processed with epitaxial lift-off due to the nature of processing the samples. Since all samples are processed in air, unlike traditional methods that use an inert atmosphere, the formation of GaAsSb-Oxide depends on the duration of their exposure to air. This sample has a smaller GaAsSb-Oxide region, resulting in a bonded region of less than 1 nm. As seen in the paper, the presence of GaAsSb-Oxide does not affect the structure or electrical performance of the GaAsSb/Si photodiodes.

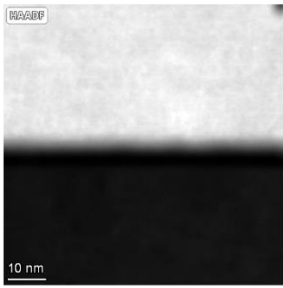
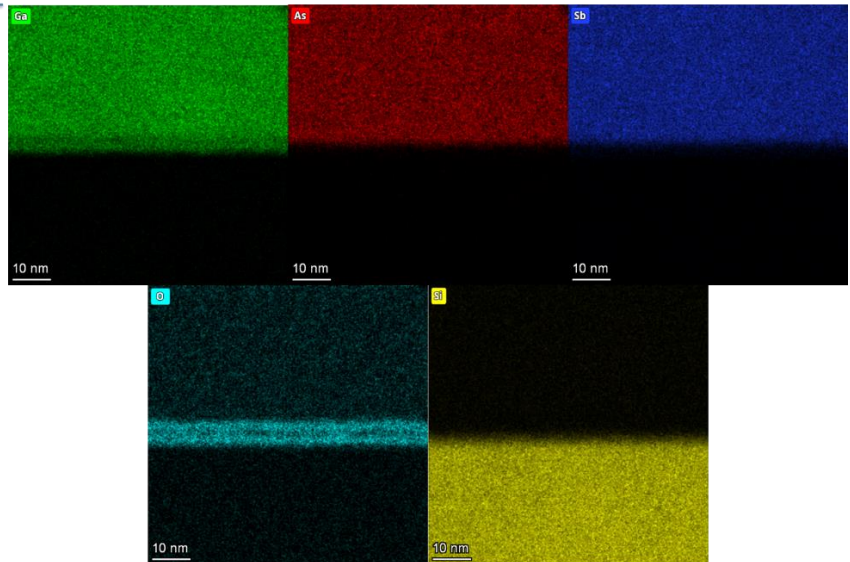

(a) GaAsSb/Si fabricated with Macro-Transfer Printing

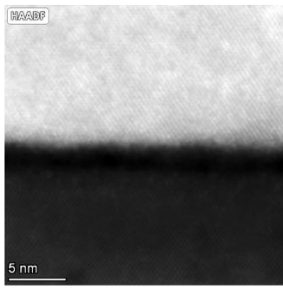
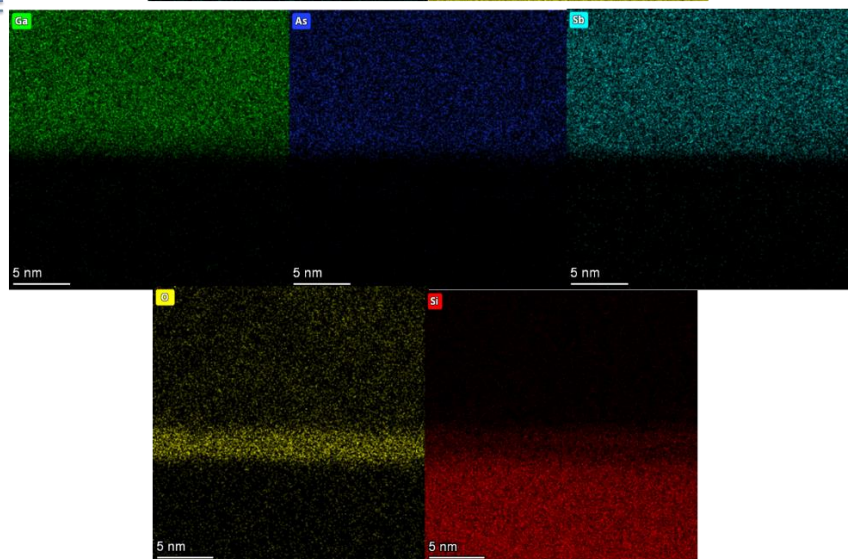

(b) GaAsSb/Si fabricated with Epitaxial Lift-Off

*Supplementary Figure 3 Individual EDS area map for GaAsSB/Si heterostructure, fabricated with (a) Macro-Transfer Printing and (b) Epitaxial Lift-off.*

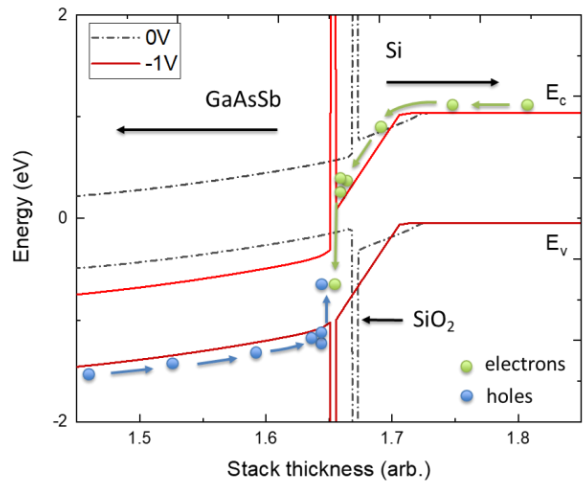

*Supplementary Figure 4 Silvaco simulation of GaAsSb/Si heterostructure with oxide at the interface, depicting hole accumulation due to large valence band off-set between GaAsSb and Si. Additionally, the simulation shows that the band-bending in the heterostructure also depends on the presence of oxide between the interfaces. Regardless, of the presence of oxide, the holes face accumulation at the interface of GaAsSb and Si due to valence bend off-set, leading to lower photocurrent in the devices.*